\documentclass[sigconf]{acmart}

\newcommand{\stitle}[1]{\noindent{\textbf{#1.}}}

\usepackage{xurl} 
\usepackage{xspace} 

\newcommand{\code}[1]{\texttt{\small\url{#1}}}

\newcommand{\hide}[1]{}

\definecolor{HighlightColor}{rgb}{0.85,0.1,0.1}
\newcommand{\revision}[1]{#1}

\usepackage{tikz}
\newcommand{\panelbadge}[2]{%
  \tikz[baseline=(char.base)]{
    \node[
    circle,
    fill=#1,
    inner sep=0pt,
    minimum size=1.3em,
    font=\normalsize\bfseries,
    text=white
    ] (char)
    {\textcolor{white}{\bfseries #2}};
  }%
}

\usepackage{xcolor}
\definecolor{panelblue}{HTML}{43B2C2}
\definecolor{panelorange}{HTML}{E69138}

\AtBeginDocument{%
  }

\setcopyright{acmlicensed}

\copyrightyear{2026}
\acmYear{2026}
\setcopyright{cc}
\setcctype{by}
\acmConference[SIGMOD Companion '26]{Companion of the International Conference on Management of Data}{May 31-June 05, 2026}{Bengaluru, India}
\acmBooktitle{Companion of the International Conference on Management of Data (SIGMOD Companion '26), May 31-June 05, 2026, Bengaluru, India}
\acmDOI{10.1145/3788853.3801596}
\acmISBN{979-8-4007-2450-3/2026/05}

\begin{document}
\title{BDIViz in Action: Interactive Curation and Benchmarking for Schema Matching Methods}

\hide{
\author{Eden Wu \ \ Christos Koutras \ \ Cl\'audio Silva \ \ Juliana Freire}
\affiliation{
  \institution{VIDA Center, New York University}
  \city{New York}
  \country{USA}
}
\email{{eden.wu,christos.koutras,csilva,juliana.freire}@nyu.edu}
}

\settopmatter{authorsperrow=4}

\author{Eden Wu}
\affiliation{
  \institution{VIDA Center, New York University}
  \city{New York}
  \country{USA}
}
\email{eden.wu@nyu.edu}

\author{Christos Koutras}
\affiliation{
  \institution{VIDA Center, New York University}
  \city{New York}
  \country{USA}
}
\email{christos.koutras@nyu.edu}

\author{Cl\'{a}udio T. Silva}
\affiliation{
  \institution{VIDA Center, New York University}
  \city{New York}
  \country{USA}
}
\email{csilva@nyu.edu}

\author{Juliana Freire}
\affiliation{
  \institution{VIDA Center, New York University}
  \city{New York}
  \country{USA}
}
\email{juliana.freire@nyu.edu}
\renewcommand{\shortauthors}{Eden Wu, Christos Koutras, Cláudio T. Silva, and Juliana Freire.}

\begin{abstract}

Schema matching remains fundamental to data integration, yet evaluating and comparing matching methods is hindered by limited benchmark diversity and lack of interactive validation frameworks. BDIViz, recently published at IEEE VIS 2025, is an interactive visualization system for schema matching with LLM-assisted validation. Given source and target datasets, BDIViz applies automatic matching methods and visualizes candidates in an interactive heatmap with hierarchical navigation, zoom, and filtering. Users validate matches directly in the heatmap and inspect ambiguous cases using coordinated views that show attribute descriptions, example values, and distributions. An LLM assistant generates structured explanations for selected candidates to support decision-making.

This demonstration showcases a new extension to BDIViz that addresses a critical need in data integration research: human-in-the-loop benchmarking and iterative matcher development. New matchers can be integrated through a standardized interface, while user validations become evolving ground truth for real-time performance evaluation. This enables benchmarking new algorithms, constructing high-quality ground-truth datasets through expert validation, and comparing matcher behavior across diverse schemas and domains.
%
We demonstrate two complementary scenarios: (i) data harmonization, where users map a large tabular dataset to a target schema with value-level inspection and LLM-generated explanations; and (ii) developer-in-the-loop benchmarking, where developers integrate custom matchers, observe performance metrics, and refine their algorithms.
%
\hide{
Our system BDIViz, recently published at IEEE VIS 2025, introduced an interactive visualization system for data integration with LLM-assisted validation. Given source and target datasets or schemas, BDIViz runs automatic schema matching methods
to generate match candidates and visualizes them in an interactive heatmap with hierarchical navigation, zoom, and filtering. Users validate matches directly in the matrix (accept/reject/annotate) and inspect ambiguous cases through coordinated views exposing attribute descriptions, example values, and value distributions. For selected candidates, an LLM-based assistant produces structured explanations using multiple matching criteria to support decision-making and conflict resolution. 
%
This demonstration showcases a new extension of BDIViz to support human-in-the-loop benchmarking and matcher development. New matchers can be integrated through a standard interface, while user-validated decisions serve as evolving ground truth for on-the-fly performance comparisons. To showcase the system's functionalities, we demonstrate two scenarios: (i) data harmonization by mapping a large tabular dataset to a given schema with value-level inspection and LLM-agent explanations; and (ii) developer-in-the-loop benchmarking, where developers integrate new matchers and observe real-time performance comparisons for a standard table-to-table matching task.}
\end{abstract}



\maketitle

\section{Introduction}

Schema matching remains a persistent bottleneck in data-centric workflows: practitioners must align attributes across heterogeneous datasets, verify that matched fields truly refer to the same concept, and reconcile value-level inconsistencies (e.g., incompatible encodings, units, or categorical conventions). While automated matching methods and systems have improved substantially \cite{koutras2021valentine, freire2025largelanguagemodel, liu2024magneto, aumueller2005coma, tao2020imi}, real deployments still require substantial human validation. This is especially true for high-stakes, terminology-dense domains, where semantic ambiguity and value-format diversity are common. At the same time, the development of matching algorithms faces a complementary challenge: progress is difficult to measure when ground truth is scarce and expensive to curate. In practice, domain experts make many 
accept/reject decisions during curation, but these signals are rarely captured in a way that can drive systematic benchmarking, diagnostics, and rapid matcher improvement.

To facilitate human validation of automated matching outputs, we recently introduced BDIViz~\cite{wu2025bdiviz}, an interactive visualization system for schema matching with LLM-assisted capabilities.\footnote{Source code and video can be accessed at: \url{https://github.com/VIDA-NYU/bdi-viz}}
Given source and target datasets or schemas, BDIViz executes multiple matchers and presents candidate correspondences in an interactive heatmap. Users validate matches directly in the matrix while coordinated views provide schema descriptions, value comparisons, and distributions. For ambiguous cases, a Harmonization Assistant generates structured rationales and supports iterative refinement.

This work extends BDIViz to support developer-in-the-loop benchmarking, by integrating a benchmarking workflow directly into the curation process. BDIViz records user-validated decisions as evolving ground truth and updates matcher evaluation views as curation progresses. Developers can 
upload a customized matcher via an in-browser code editor, run it for the current task, and observe how it compares against existing matchers.

The demonstration engages attendees in two interactive scenarios corresponding to distinct user roles. In the first, participants act as data curators, performing end-to-end harmonization of a large biomedical dataset to a standard schema. They navigate the scalable heatmap interface, validate matches with value-level inspection, and use the LLM assistant to resolve ambiguous correspondences. In the second scenario, they take on the developer role, implementing a new matcher via the in-browser editor, running it on the current task, and analyzing its performance against existing matchers through benchmarking views that expose failure modes and suggest improvements. Both scenarios support export of mapping specifications and curated ground truth for offline reuse.
\section{System Overview}
\label{sec:system_overview}

\begin{figure}
  \centering
  \includegraphics[width=\linewidth]{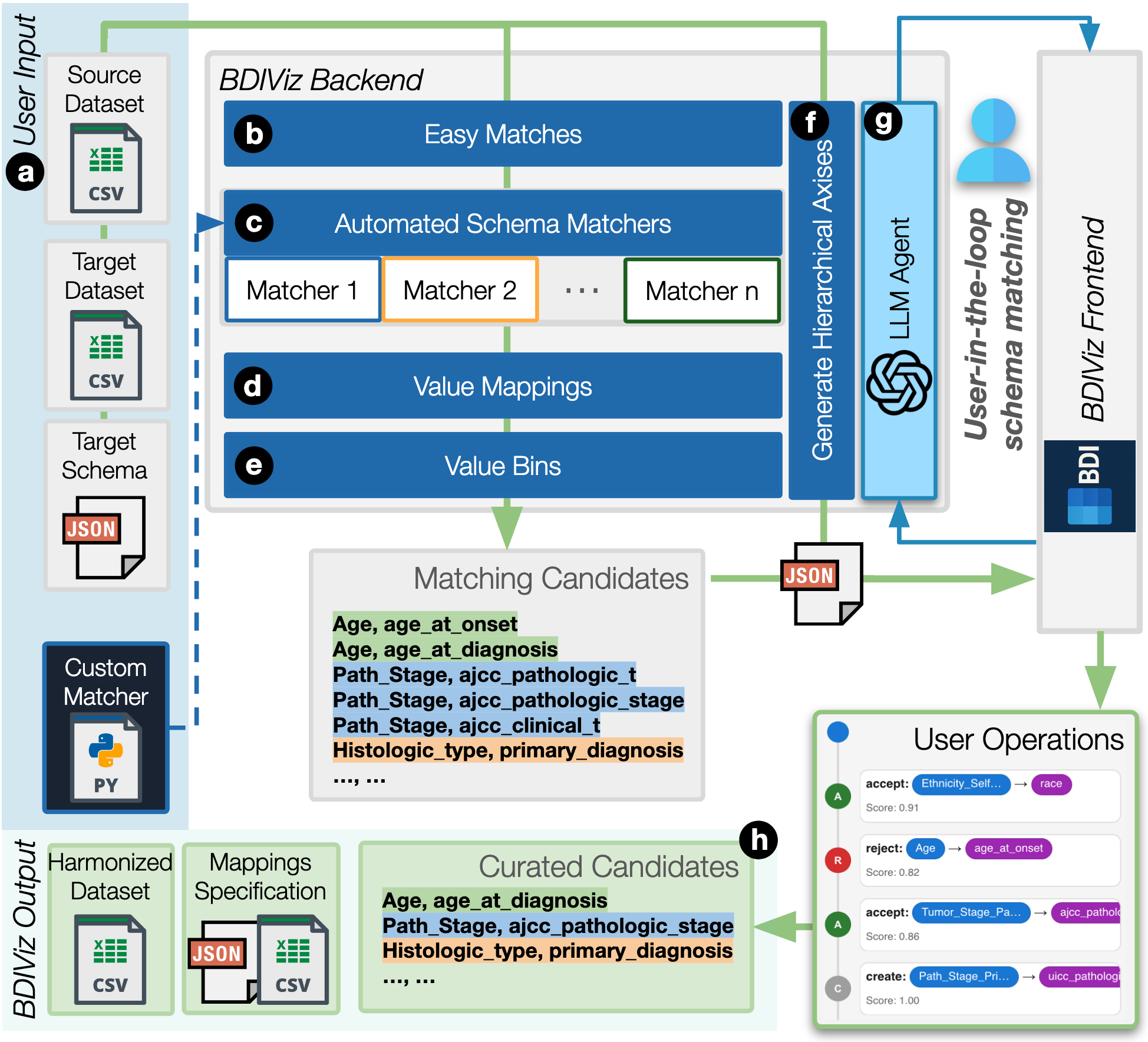}
  \vspace{-.7cm}
  \caption{BDIViz system overview}
  \label{fig:system}
  \vspace{-.7cm}
\end{figure}

In this section, we provide an overview of BDIViz, as illustrated in \autoref{fig:system}, focusing on key components and functionality.



\subsection{Matching Pipeline and Live Ground Truth}
\label{sec:backend}

The backend converts raw inputs \panelbadge{black}{a} into a set of persistent artifacts consumed by the frontend: inferred ontologies, ranked match candidates with per-matcher scores, value comparisons and distribution bins, provenance logs, and evaluation summaries. To keep the interface responsive, long-running computations run asynchronously in two Celery worker pools dedicated to ontology inference and candidate generation and value profiling.

\stitle{Ontology inference}
When the source or target schema is not provided, the ontology worker infers attribute properties and semantic groupings of attributes for both the source and target \panelbadge{black}{f}. The resulting ontology is stored in JSON
and is later used to drive hierarchy-aware filtering and ontology-based search in the client. 

\stitle{Candidate generation}
Candidate generation proceeds in three steps:
(i) Easy match detection identifies high-confidence correspondences using fuzzy similarity and marks them for auto-acceptance to reduce review load \panelbadge{black}{b}.
(ii) Matcher-agnostic ensemble execution runs an extensible ensemble of matchers (heuristic, embedding-based, and SLM/LLM-based)~\cite{koutras2021valentine,lopez2026bdikit,liu2024magneto} and aggregates their outputs into ranked candidates per source attribute. The ensemble is matcher-agnostic: developers can register additional matchers at runtime through a standardized Python interface, enabling integration of custom algorithms without system modifications. \revision{For large schemas or computationally expensive matchers, this stage may require significant processing time; however, it can be executed offline and its results reused via cached artifacts.}
The backend executes the new matcher within the same pipeline and materializes its scores into the candidate artifacts \panelbadge{black}{c}. 
BDIViz maintains per-matcher weights that are updated from user accept/reject feedback, allowing rankings to adapt to task-specific semantics within a session.
(iii) Value profiling and mapping artifacts compute summaries to support interactive inspection, including frequency counts for categorical attributes and binned histograms for numeric attributes \panelbadge{black}{e}. The backend also generates a one-to-one value-mapping proposal using fuzzy matching on unique values \panelbadge{black}{d}, enabling the Value Comparison preview and the dataframe-level Value Wrangler view.

\stitle{Provenance and export}
As users interact with the frontend (e.g., accepting or rejecting candidates, editing value mappings), BDIViz appends each operation to a provenance timeline stored with the session artifacts. Accepted correspondences are recorded as live ground truth and are immediately available to the developer loop for in-situ evaluation views and regression checks. BDIViz exports (i) the harmonized dataset and (ii) mapping specifications at both the attribute and value levels (CSV/JSON), enabling reproducible reuse and re-import \panelbadge{black}{h}.

\subsection{Agentic Validation Assistant}
\label{sec:agent}

BDIViz includes an agentic copilot that operates over backend state and artifacts. The assistant supports two functions: grounded explanation generation for selected candidates and tool-based workflow actions that modify the current task.

\stitle{Candidate explanations from multiple criteria}
When a user selects a candidate correspondence, the assistant generates a structured explanation based on multiple matching criteria and synthesizes them into an overall diagnosis (likely match vs.\ mismatch). The criteria cover complementary signals, including name similarity, token patterns, semantic meaning from schema descriptions, value-level compatibility, distributional patterns, historical mappings, and domain knowledge. \revision{LLM-based assistance introduces additional latency relative to algorithmic matching, but it remains outside the critical path of candidate generation: users can browse and curate immediately, while the assistant is invoked selectively for ambiguous matches.
In practice, single-candidate responses typically arrive within a few seconds, with higher latency for larger retrieved context or heavier server load.}

\newcommand{\schema}{schema\xspace}

\stitle{Supervisor routing with vector-store memory}
The assistant uses a multi-agent supervisor architecture with ChromaDB-backed vector store for task-specific context retrieval.
The vector store indexes 
(i) \schema information,
and (ii) user-in-the-loop operations, enabling retrieval of task-specific context. A supervisor agent routes requests to specialized agent(s): 
a Schema Agent that retrieves additional candidates from the target ontology;
a Candidate Agent that appends/prunes/accepts/rejects/edits candidates; a Task Agent that reruns matching with filters and adds/deletes matchers; and a Value Agent that proposes value mappings, dictionary lookups, and simple numeric transforms. Agent actions update the same persisted artifacts as direct UI operations, ensuring consistency between curation, provenance, and live benchmarking.

\subsection{Interactive Curation and Benchmarking}

\label{sec:frontend}

The frontend provides a coordinated workspace with views for both curation and benchmarking. The original curation views include (i) an interactive heatmap for triaging large match spaces, (ii) drill-down panels for value evidence and mapping previews, and (iii) an explanation interface for LLM-generated rationales. A principal contribution of this work is the addition of developer-oriented benchmarking views: real-time matcher performance dashboards, consensus analysis across multiple matchers, and diagnostic visualizations computed over the live, evolving ground truth. Section~\ref{sec:demo_scenario_2} presents detailed interaction sequences.

\section{Demonstration Scenarios}
\label{sec:demo}

We describe the main functionalities of our frontend interface and discuss how the users can engage with them, through two distinct representantive scenarios. The first one focuses on harmonizing a large tabular dataset with respect to a given schema, while the second one investigates how our system can be used for evaluating a matching method based on manual curation.

\subsection{Scenario 1: Biomedical Data Harmonization}
\label{sec:demo_scenario_1}

\begin{figure*}
  \centering
  \includegraphics[width=\linewidth]{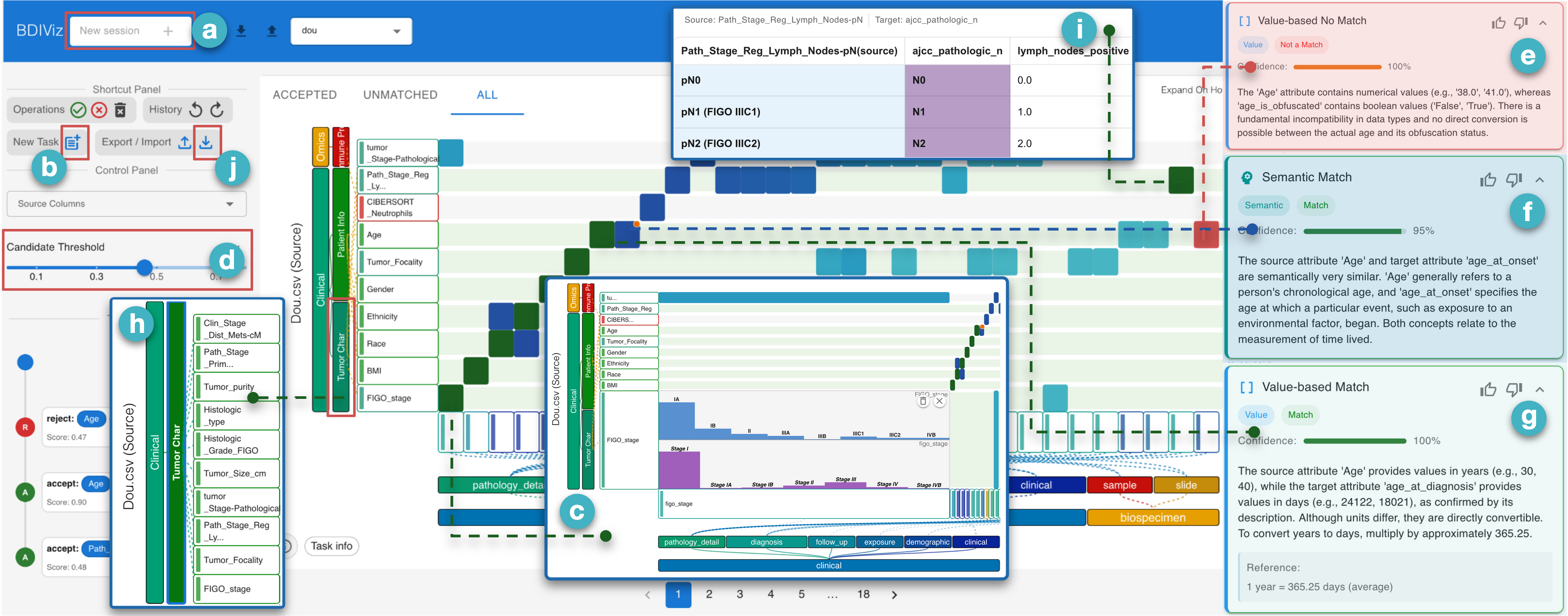}
  \vspace{-18pt}
  \caption{Demonstration Scenario 1: BDIViz offers a set of interactive and LLM-assisted visual tools to users towards data harmonization.}
  \label{fig:case_1}
  \vspace{-10pt}
\end{figure*}

In this scenario, the user acts as a domain curator and uses BDIViz to harmonize a CPTAC~\footnote{\url{https://gdc.cancer.gov/about-gdc/contributed-genomic-data-cancer-research/clinical-proteomic-tumor-analysis-consortium-cptac}} dataset from an endometrial cancer study to the GDC\footnote{\url{https://portal.gdc.cancer.gov/}} schema. The source dataset contains 179 attributes, which must be aligned to a target model with 736 attributes \cite{santos2025gdcsm}.

\stitle{Step 1: Task setup and automatic matching}
The user starts a new session \panelbadge{panelblue}{a} and creates a task by uploading the source dataset and target schema \panelbadge{panelblue}{b}. BDIViz immediately runs the matcher ensemble in the backend, while the UI reports progress and streams intermediate results into the visual interface.

\stitle{Step 2: Triage at scale in the matching matrix}
BDIViz renders the candidate space as an interactive heatmap matrix \panelbadge{panelblue}{c}, where rows are source attributes, columns are target attributes, and cell intensity encodes aggregated matcher confidence. Users can accept, reject, and annotate candidates directly in the matrix, enabling rapid triage across potentially thousands of candidate correspondences. High-confidence candidates are automatically accepted, and expanding a cell reveals value-distribution comparisons as immediate supporting evidence. Coordinated panels update synchronously with schema metadata and value-level summaries, allowing users to quickly confirm routine mappings and concentrate on ambiguous regions.


\stitle{Step 3: Resolving an ambiguous attribute with grounded explanations}
The user inspects a flagged attribute for which no candidate exceeds the confidence threshold. By adjusting the similarity cutoff, additional candidates become visible for review \panelbadge{panelblue}{d}. For one candidate, expanding the cell reveals incompatible value domains—numerical values on the source side versus boolean values on the target side. The Harmonization Assistant’s explanation panel corroborates this mismatch using instance-level evidence, and the user rejects the candidate \panelbadge{panelblue}{e}.
For the remaining plausible candidates, the explanation panel provides an overall diagnosis together with criterion-specific evidence (e.g., semantic meaning, naming patterns, and value compatibility). In one case, missing target values prevent a definitive decision, and the user records a note to flag the candidate for follow-up \panelbadge{panelblue}{f}. In another case, the assistant identifies a systematic unit mismatch and recommends an appropriate transformation, enabling the user to accept the correspondence with a justified conversion \panelbadge{panelblue}{g}.

\stitle{Step 4: Category drill-down and value mapping checks}
To further narrow the search space, the user filters the matrix using hierarchy-aware axes to focus on a specific semantic category \panelbadge{panelblue}{h}. Within this subset, a strong candidate is identified. The Value Comparison view proposes an initial one-to-one alignment over unique values \panelbadge{panelblue}{i}, while the assistant highlights a consistent encoding difference and suggests a normalization rule. The user accepts the match and verifies the transformation using both the unique-value preview and the dataframe-level Value Wrangler view to catch row-level edge cases before finalizing the mapping.

\stitle{Step 5: Provenance and export}
Throughout the workflow, BDIViz records a timeline of user actions, enabling review and collaboration. At completion, the user exports (i) the harmonized dataset and (ii) a complete mapping specification capturing attribute correspondences and value mappings, so the integration decisions are reproducible and reusable in downstream analyses \panelbadge{panelblue}{j}.

\vspace{-8px}
\subsection{Scenario 2: In-situ Matcher Evaluation}
\label{sec:demo_scenario_2}

\begin{figure}[t]
  \centering
  \includegraphics[width=.9\linewidth]{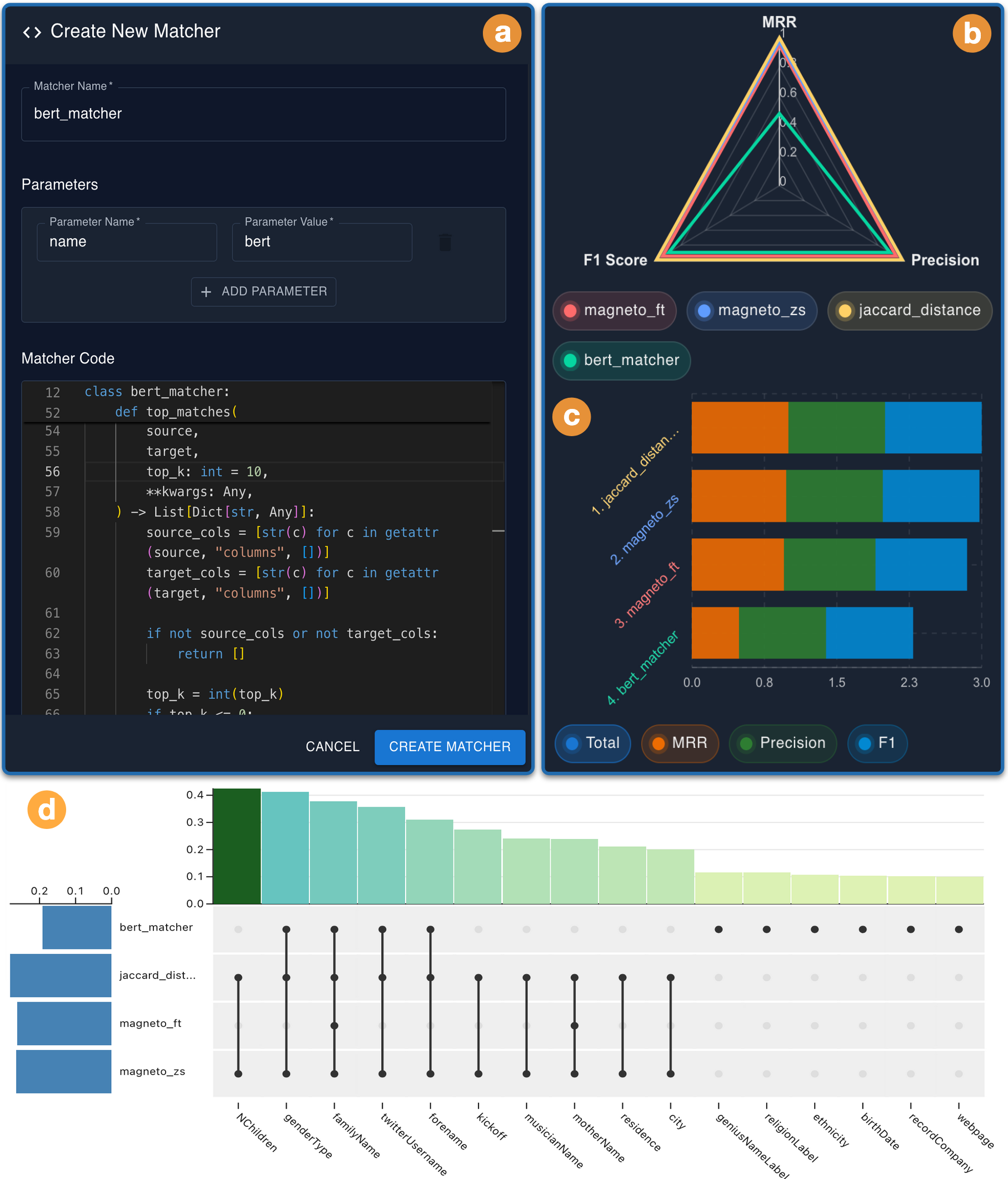}
  \vspace{-18pt}
  \caption{Demonstration Scenario 2: BDIViz enables developers to upload, evaluate and compare their matching methods through dedicated panels.}
  \label{fig:case_2}
  \vspace{-.7cm}
\end{figure}

In this scenario, the user acts as a matcher developer and evaluates a BERT encoder as a schema matcher on the manually curated WikiData benchmark~\cite{koutras2021valentine}.

\stitle{Step 1: Task initialization and developer mode}
The user creates a new session (\code{wikidata}) and initializes a task by uploading the WikiData source and target CSVs. After the candidate generation completes, the user toggles Developer Mode to expose matcher-oriented controls and evaluation views.

\stitle{Step 2: Adding a new matcher at runtime}
From the shortcut panel, the user selects Create New Matcher, which opens an in-browser code editor \panelbadge{panelorange}{a}. The user pastes a minimal HuggingFace-based BERT matcher that implements the required \code{top\_matches} interface by embedding column names, computing cosine similarity, and returning the top-$k$ target candidates per source attribute. Submitting the code registers the matcher in the backend and immediately runs it against the active task; the matcher then appears in the Matcher Control and Analytics panels alongside the default matchers.

\stitle{Step 3: In-situ curation with live performance updates}
As the user validates candidates in the matrix, evaluation metrics update continuously, enabling developers to observe how matcher quality evolves as more task-specific evidence is accrued. Early in the session—when ground truth consists only of auto-accepted trivial matches—all matchers achieve near-perfect metrics, obscuring real performance differences.

\stitle{Step 4: Diagnosing ranking quality and matcher agreement}
As the user continues curating and the live ground truth grows, the Matcher Analytics radar plot—which summarizes performance metrics (e.g., Precision, F1, and MRR) over the evolving live ground truth—updates to reflect the evolving performance landscape \panelbadge{panelorange}{b}. The user then inspects the Ranked Breakdown view \panelbadge{panelorange}{c}, which reveals a characteristic failure mode: while BERT can achieve reasonable Precision and F1, its MRR is substantially lower (e.g., $\approx 0.49$), indicating that correct matches are often ranked far down the candidate list and are therefore less useful in an interactive top-$k$ review workflow.

To examine agreement patterns, the user uses the UpSet-style consensus view, which shows limited overlap between BERT and the best-performing matchers on the accepted correspondences \panelbadge{panelorange}{d}. This lack of consensus further suggests that BERT's high-scoring candidates are not consistently aligned with the strongest lexical signals for this benchmark.

\stitle{Step 5: Benchmark-driven conclusion}
By the end of curation, BDIViz provides a final, task-specific performance summary based entirely on the curated correspondences. In this scenario, a baseline that leverages syntactic similarities (e.g., Jaccard distance) achieves near-perfect scores on this benchmark. This outcome indicates that the WikiData task is dominated by straightforward lexical alignment, where name similarity alone is sufficient and more complex pretrained encoders do not provide additional benefit. 
This scenario demonstrates BDIViz's unique capability for immediate, task-grounded matcher evaluation: developers inject custom matchers into active curation sessions and observe performance metrics evolve in real-time as domain experts validate correspondences, enabling rapid diagnostic feedback.

\section*{Acknowledgments}
This work was supported in part by DARPA ASKEM (HR0011262087), ARPA-H BDF, and NSF (OAC-2411221). The views, opinions, and findings expressed are those of the authors and should not be interpreted as representing the views or policies of these agencies.

\bibliographystyle{ACM-Reference-Format}
\bibliography{bdiviz-sigconf}



\end{document}